\documentstyle[emulateapj,psfig]{article}

\def\rosat{{\sl ROSAT}}

\def\chandra{{\sl Chandra}}
\def\Chandra{{\sl Chandra}}

\newcommand\psr{\hbox{PSR~B0540$-$69}}
\newcommand\snr{\hbox{SNR~B0540$-$69}}

\slugcomment{To appear in the Astrophysics Journal Letters}

\lefthead{Gotthelf \& Wang}
\righthead{X-ray nebula around \psr}

\begin{document}

\title{A spatially resolved plerionic X-ray nebula around \psr}

\author{E. V. Gotthelf$^1$ and Q. Daniel Wang$^2$}
\altaffiltext{1}{Columbia Astrophysics Laboratory, Columbia University, 550 
West 120$^{th}$ Street, New York, NY 10027, USA; evg@astro.columbia.edu}
\altaffiltext{2}{Astronomy Department, University of Massachusetts, B-524 
LGRT, Amherst, MA 01003, USA; wqd@astro.umass.edu}

\begin{abstract}

       We present a high resolution \Chandra\ X-ray observation of
\psr, the Crab-like 50 msec pulsar in the Large Magellanic Cloud. We
use phase-resolved imaging to decompose the extended X-ray emission,
as expected of a synchrotron nebula, from the point-like emission of
the pulsar.  The image of the pulsed X-ray emission shows a
well-defined point-spread function of the observation, while the
resolved nebula has a morphology and size remarkably similar to the
Crab nebula, including evidence for a jet-like feature from \psr.  The
patchy outer shell, which most likely represents the expanding
blast-wave of the supernova, is reminiscent of that seen in
radio. Based on morphology, size, and energetics there can be little
doubt that \snr\ is an analogous system to the Crab but located in our
neighboring galaxy.

\end{abstract}

\section {Introduction}

The X-ray-bright 50 ms pulsar \psr\ was discovered in the N158A nebula
of the Large Magellanic Cloud (LMC) and has long been compared to the
Crab pulsar (Seward, Harnden, \& Helfand 1984).  Based on its timing
and spectral properties, the two rotation powered pulsars are very
similar with a spin period of 50 vs. 33 ms and a spin-down rate of
$4.8 \times 10^{-13}$ vs. $4.0 \times 10^{-13}$ s/s for \psr\ and the
Crab, respectively. From these quantities, assuming the standard
magnetic dipole pulsar model, one can infer a characteristic age (1.7
vs. 1.2 kyr), spin-down energy ($2.0 \times 10^{38}$ vs. $6.5 \times
10^{38}$ ergs s$^{-1}$), and surface magnetic field strength ($5.0
\times 10^{12}$ vs. $3.8 \times 10^{12}$ G) for \psr\ and the
Crab, respectively.  This similarity suggests that \psr\ could be
accompanied by a ``plerion'', a pulsar driven wind nebula (Weiler \&\
Sramek 1988), reminiscent of that seen for the Crab.

Indeed, there are several lines of evidence indicating the presence of
a plerion in the vicinity of \psr. Chanan, Helfand, \& Reynolds (1984)
detected a polarized optical nebula of half-power diameter $\sim
4^{\prime\prime}$ around the pulsar. This apparent synchrotron nebula
was also resolved (though barely) in a radio image presented by
Manchester et al.  (1993) and in a \rosat\ high resolution imager
observation by Seward \& Harnden (1994). Furthermore, the overall
X-ray spectrum of \psr\ and its remnant (\snr) is well characterized by a
power law, as expected if the emission is predominantly non-thermal
(Clark et al. 1982; Wang \& Gotthelf 1998a).

The \rosat\ observation also revealed a faint X-ray emitting shell,
$\sim 15$~pc in size surrounding the pulsar. This shell contributes
less than $ 20\%$ to the total luminosity of $\sim 1.0 \times 10^{37}
{\rm~ergs~s^{-1}}$ in the \rosat\ 0.1-2~keV band and is likely the SNR
associated with the pulsar (Seward \& Harden 1994). However, no
evidence has yet been found for a similar X-ray-emitting shell or a
shell-like SNR around the Crab (e.g. see discussion in Jones et
al. 1998).

In this Letter, we report new results on \snr\ based on a recent
observation acquired with the \Chandra\ High Resolution Camera.  This
observation enables us for the first time to distinguish morphological
details of the nebula around \psr. We analyze phase dependent images
and resolve the expected plerion-like nebula from the point-like
pulsar emission. This allows us to identify features similar to those
seen from the Crab nebula; we present morphological evidence for a
torus of X-ray emission, which most likely represents shocked pulsar wind 
materials and a likely X-ray jet emanating from the pulsar. We discuss the
implications of the results in the context of the pulsar-nebula
system. Throughout the paper we adopt a distance of $51$~kpc for the
LMC.

\section {Observation}

The \Chandra\ observatory (Weisskopf et al. 1996) observed \snr\ on
Aug 31 1999 as part of the initial calibration of the High Resolution
Camera (HRC; Murray et al. 1997). A total of 19.4 ksec of data were
collected during a portion of the orbit which avoided regions of high
background contamination such as from bright Earth and radiation belt
passages.  The remnant was centered on the on-axis position of the HRC
where the point-spread function (PSF) has a half power radius (the
radius enclosing 50\% of total source counts) of $\sim 0\farcs5$.
Time-tagged photons were acquired with 15.6 $\mu$s precision, and the
arrival times were corrected to the solar system barycenter using a beta
version of {\tt AXBARY} available from the \Chandra\ X-ray Center FTP
site (A. Rots 1999, private communication). While the detector is sensitive
to X-rays over a $0.1-10.0$ keV range, there is essentially no energy
information available.  We analyzed event data calibrated by the
initial processing and dated 1999 September 12, which was made
available through the \Chandra\ public archive. In addition to the
standard processing, the event data was further filtered to reduce the
instrumental background and to remove ``ghost image'' artifacts using a
beta version of {HRC\_SCREEN} (S. Murray 2000, private
communication). We extracted $1024 \times 1024$ pixel images centered
on the pulsar rebinned by a factor of two from the native HRC pixel
size of 0\farcs13175 per side.

\section {Results}

A global view of N158A and its environment as seen by the \Chandra\
HRC is shown in Figure~$1$ (contours) and Figure~$2$ (greyscale).  The
large-scale X-ray enhancement on scales up to $\sim 1^\prime$ is the
previously resolved shell-like emission (Seward \& Harnden 1994).  In
fact, the X-ray and radio emission together outlines a nearly circular
morphology around \psr. Clearly, the shell represents the blastwave of
\snr. The X-ray intensity distribution within the remnant appears
rather patchy. While the southwest X-ray enhancement is a good tracer
of the radio and optical emission peaks, there is no general
correlation between fainter radio and X-ray features.

The superb spatial resolution of the \Chandra\ observation further
allows for a close-up of the immediate vicinity of \psr\ (Fig.~$2b$).
The presence of a diffuse plerion-like nebula around the pulsar is
apparent. To decompose the nebula emission from the pulsar
contribution, we conducted phase-resolved image analysis. This enables
us to estimate the local PSF based on the pulsed, point-like emission
from the pulsar and to quantify the extended, unpulsed nebula
radiation.

First, we must determine the pulse period at the current epoch. We
constructed a periodigram around a narrow range of periods centered on
the expected period $\pm 0.1$ ms, sampled in increments of $0.05(P^2/T)$,
were $T$ is the observation duration, and $P$ is the test
period.  For each trial period, we folded photons extracted from a
1\farcs0 aperture centered on the bootstrapped pulsar position (see
below) into 20 phase bins and computed the $\chi^2$ of the resultant
profile. We find a highly significant signal ($> 56 \sigma$) at $P =
50.508132(6)$ ms at Epoch 51421.630741 MJD; the uncertainty is
estimated according to the method of Leahy (1987). We have assumed a
period derivative of $\dot P = 4.789342 \times 10^{-13}$ for the data
epoch (Deeter et al. 1999). In Figure~$3$ we display the resultant
light-curve folded at the peak period which, as expected, is roughly
sinusoidal and modulated with a $\sim 40\%$ pulse fraction (defined as
the amplitude divided by the mean).

Next, we defined two regions in the phase space, on- and off-pulse, by
selecting eight adjacent phase bins corresponding to the peak and
trough of the pulse profile. The on-pulse image with the off-pulse
image subtracted is presented in Figure~$4b$. This image
reproduced the expected PSF with no evidence of asymmetric deviations,
as might be caused by poor aspect reconstruction, like that typically
found for \rosat\ images. Figure~$5$ presents average radial intensity
distributions around the centroid of the point-like source.

By subtracting the normalized pulsar image (Fig.~$4b$), we are able to
construct an image of the nebula emission (Fig.~$4c$) alone.  The
subtracted image is scaled to compensated for both the relative phase
coverage and for a 21\% unpulsed emission contribution, estimated by
minimize a pointlike contribution at the pulsar position of the nebula
image. As shown in Figure~$4c$, the extended emission is distinctly
different from the pointlike image of the pulsed emission from the
pulsar. The primarily feature is the NE-SW elongated feature, which is
morphologically symmetric relative to the pulsar and extends about
$\sim 2\farcs5$ on both sides of the pulsar.  However, the observed
emission on the SW side appears twice as strong compared to the NE
side, with an average intensity of $\sim 7.5 \times 10^{-2}
{\rm~counts~s^{-1}~arcsec^{-2}}$.  Because the central core of the
distribution is significantly brighter than the extended features and
the subtraction of the pulsar contribution is somewhat arbitrary, the
exact intensity distribution is uncertain.

There is also marginal evidence for a jet-like feature emanating from
the pulsar. This emission, most apparent in the NW and extending about
$3^{\prime\prime}$, is nearly perpendicular to the NE-SW elongated
nebula and is slightly bent toward North. The integrated emission of
the jet-like feature is roughly $\sim 3.1 \times 10^{-2}
{\rm~counts~s^{-1}}$. The configuration of the jet feature relative to
the nebula is remarkably similar to that of the Crab nebula as seen by
ROSAT and which is now clearly resolved with Chandra (see Chandra
publicity photo).

In short, the X-ray emission can be decomposed into three major
morphological components: a point-like source, the surrounding nebula
which shows evidence for a jet feature, and a patchy supernova remnant
shell.

\begin{deluxetable}{lcc}
\tablewidth{0pt}
\tablecaption{HRC Spatial Components of \snr
\label{tbl-1}}
\tablehead{
\colhead{Component$^a$} & \colhead{Count Rate$^b$} &  \colhead{Size and Shape}
}
\startdata
Pulsar$^c$ &   & \nl
\quad pulsed &  0.14  & point-like \nl
\quad unpulsed &  0.18  & '' \nl
Nebula     &  0.8  & $5^{\prime\prime} \times 3^{\prime\prime}$ NE-SW \nl
Jet        &  0.03 & $3^{\prime\prime}$ long SE-NW \nl
SNR Shell  &  0.2  & $\sim 1^{\prime}$ diameter\nl
\enddata
\tablenotetext{a}{See \S 3; $^b$Count/s; $^c$X-ray emission
from a $2^{\prime\prime}$ radius aperture around the pulsar.}
\end{deluxetable}

\section {Discussions}

	A comparison between the X-ray-emitting nebula around \psr\
and the Crab nebula (see Chandra publicity photo\footnote{Available at
\hbox{\tt http://xrtpub.harvard.edu/photo/0052/0052\_hand.html}}) is
very informative.  The Crab nebula image shows a torus of
X-ray-emitting loops, which most likely represents shocked pulsar wind
materials consisting of magnetic waves and ultra-relativistic
particles. Also clear are the two jets of X-ray-emitting material
emanating from the Crab pulsar, in the direction perpendicular to the
major axis of the torus. We speculate that the nebula around \psr\ has
a similar structure. In fact, at the spatial resolution of Chandra at
the LMC, the size, morphology, and surface intensity of the two
nebulae are all remarkably similar (Fig. $2b$).

	Assuming a power law spectrum for the X-ray emission from the
\snr\ nebula of photon index 2.0 and N$_H = 4 \times 10^{21}
\rm{cm}^2$ (Finley et al. 1993, Wang \& Gotthelf 1998a), the
conversion between the count rate and the energy flux in the standard
1.0-10~keV band is $\sim 1 \times 10^{-10} {\rm~ergs~s^{-1}~cm^{-2}
/counts~s^{-1}}$. The corresponding total luminosity of the nebula is
$\sim 2.7 \times 10^{37} {\rm~ergs~s^{-1}}$, which is 13\% of the
spin-down energy of \psr. The fraction is again similar to that
of the Crab.
	
	N157B (PSR J0537$-$6910) is the only other LMC SNR with a
detected pulsar (16~ms) and also shows both an extended (resolved by
\rosat\ HRI), non-thermal nebula and a partial X-ray-emitting shell
(Wang \& Gotthelf 1998a; 1998b). 
The upcoming \chandra\ observations
will make a detailed comparison between these two young Crab-like 
SNRs possible. 

\acknowledgements

We gratefully acknowledge the Chandra team for making available the
public data used herein. In particular we thank A. Rots and S. Murray
for kindly making available their beta software. We thank U. Hwang for
pointing out an instrumental artifact (``ghost image'') in the original
HRC image of SNR B0540-69. We thank Dick Manchester for sending us the
radio image and Fernando Camilo, David Helfand, and Jules Halpern for
carefully reading the manuscript.  This work was funded in part by
NASA LTSA grants NAG5-7935 (E.V.G.) and NAG5-6413 (Q.D.W.). This is
contribution \#690 of the Columbia Astrophysics Laboratory.

\begin{figure} 
\centerline{ {\hfil\hfil
\psfig{figure=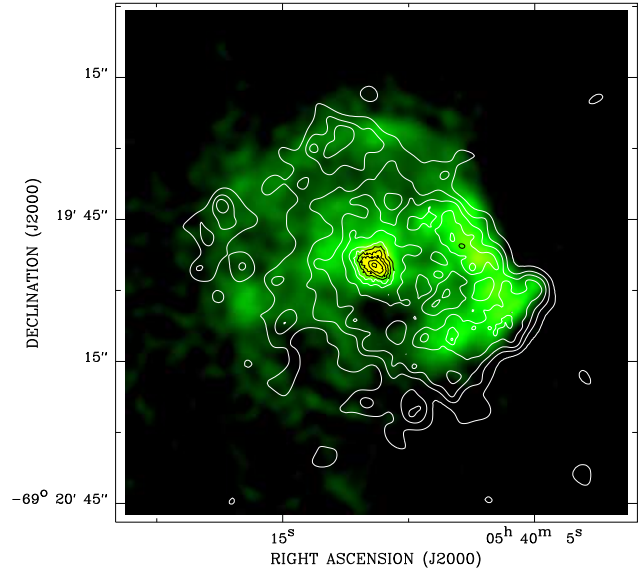,height=3.0in,angle=90, clip=}
\hfil\hfil} }
\caption{The overall X-ray and radio view of the region of SNR~0540-69
containing the central 50 ms pulsar \psr. The contours denote the
Chandra HRC X-ray intensity distribution; the levels are
0.001, 0.005, 0.008, 0.013, 0.03, 0.1, 0.3, 1., 3, 10, and 30 counts s$^{-1}$
arcsec$^{-2}$. The greyscale map shows the 6 cm radio emission which
traces the approximately circular remnant.  The radio map is from
Manchester et al. (1993) and has a beam size with half power width of
$\sim 5"$.
\label{fig1}}
\end{figure}

\begin{figure} 
\centerline{ {\hfil\hfil
\psfig{figure=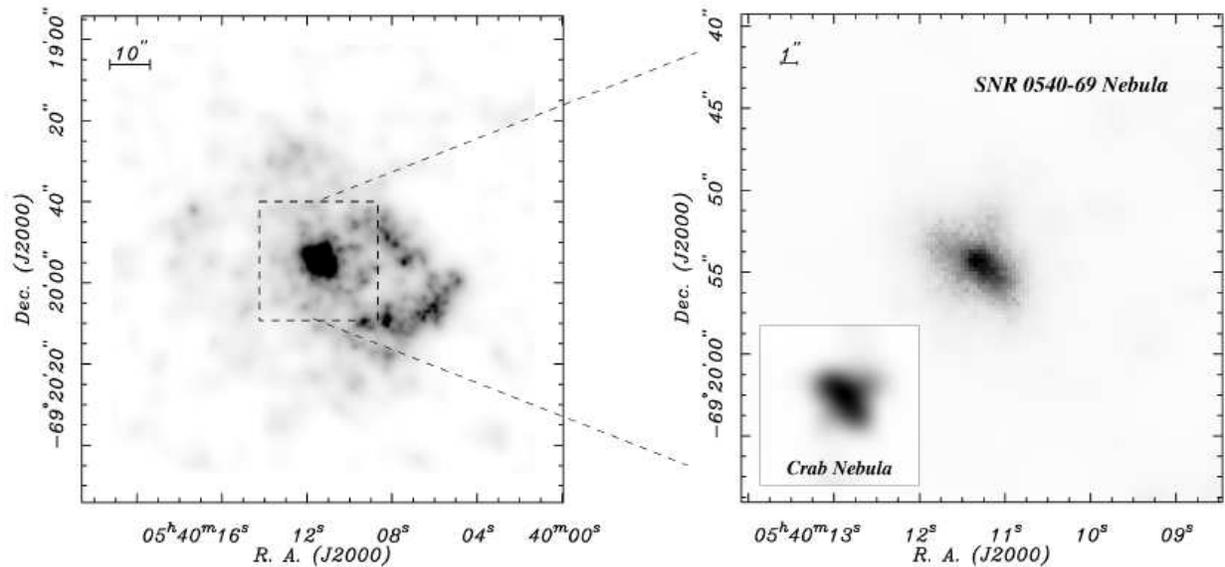,height=3.0in,angle=270,clip=} 
\hfil\hfil} }
\caption{X-ray intensity distribution around the region containing the
50 ms pulsar \psr. Left Panel) the Chandra HRC broad energy-band image
centered on the pulsar. The image has been adaptively smoothed using
a minimum signal-to-noise of 6 ratio and the intensity scale chosen to
highlight the diffuse SNR emission surrounding the bright pulsar
nebula, which is fully saturated in this image. The central box
delineates the region enlarged and displayed in the right panel. Right
Panel) Close up image around the pulsar, displayed scaled by the
square root of intensity. Insert) Crab nebula image, placed at the
distance of the LMC (assuming a distance to the Crab of 2 kpc) and
blurred to the HRC resolution. Notice the similar size, shape, and
overall brightness morphology compared to \psr.
\label{fig2}}
\end{figure}

\begin{figure} 
\centerline{ {\hfil\hfil
\psfig{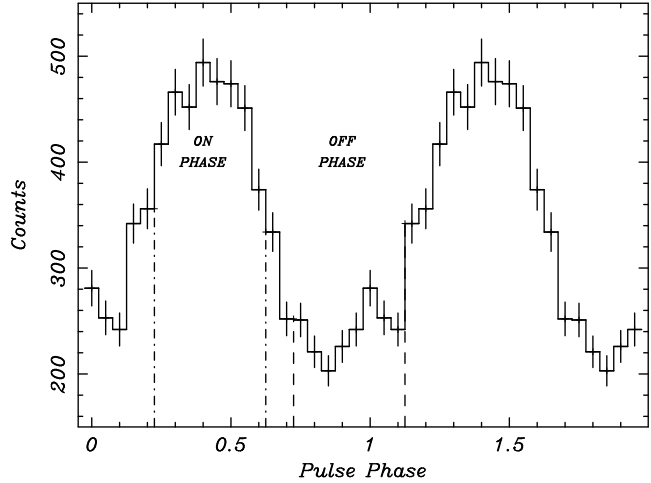}
\hfil\hfil} }
\caption{HRC light curve ($0.4 - 10$ keV) for \psr\ folded at the
ephemeris given in the text. Two periods are shown for clarity.  The two sets 
of vertical
bars denote the two phase regions (on/off) used to isolate the nebula from the
pulsar emission. The light curve has been extracted from a $1^{\prime\prime}$
radius aperture centered on the pulsar.
\label{fig3}}
\end{figure}

\bigskip\bigskip

\begin{figure} 
\centerline{ {\hfil\hfil
\psfig{figure=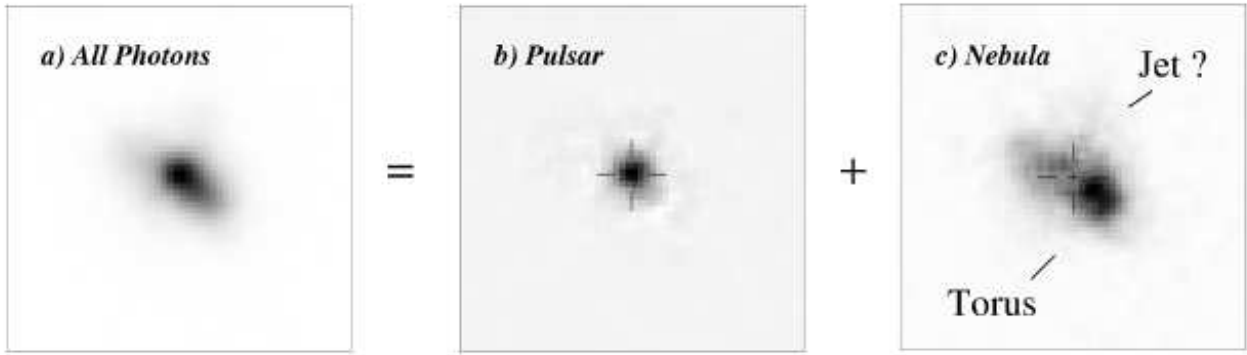,height=2.0in,angle=270, clip=}
\hfil\hfil} }
\caption{Phase-subtracted X-ray intensity maps of \psr.  Left Panel)
Before phase-subtraction, image of the pulsar and nebula (all photons),
same as the right panel of figure 2. Middle Panel) the pulsar image
containing the pulsed photons only. This is in excellent
representation of the HRC point-spread function, consistent with the
ground calibration. Right Panel) same region after subtracting the
pulsar's contribution to the total flux (see text for details) -- this
provides a good estimate of the nebula emission surrounding the
pulsar. The cross marks the pulsar's centroid. The three maps are
identically sized and linearly scaled in intensity.
\label{fig4}}
\end{figure}

\begin{figure} 
\centerline{ {\hfil\hfil
\psfig{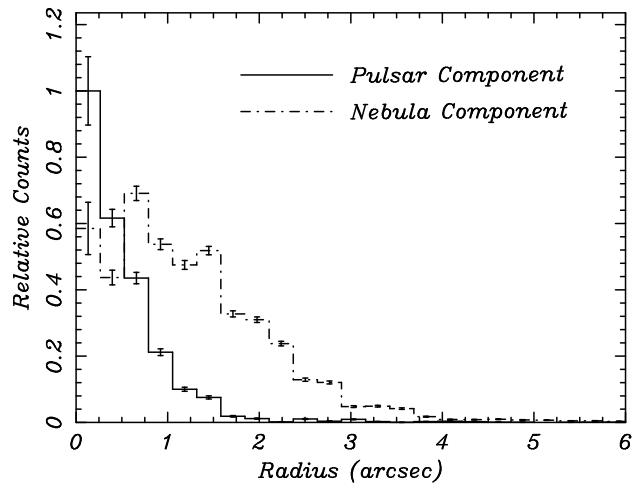}
\hfil\hfil} }
\caption{HRC radial intensity distributions around the point-like
emission peak of \psr. The pulsar radial profile (solid line) is
consistent with the HRC point-spread function. The nebula profile
(dotted line) is significantly extended, the enhancement between
$0.1-1.5^{\prime\prime}$ is due to the relatively bright SW nebula
region.
\label{fig5}}
\end{figure}


\begin{references} 
\reference{} Caraveo, P., Mignani, R., \& Bignami, G. B. 1998, in Memorie 
della Societa Astronomia Italiana, Vol. 69, p.1061
\reference{} Chanan, G. A., Helfand, D. J., \& Reynolds, S. P. 1984 ApJ, 287, 
L23
\reference{} Clark, D. H., Tuohy, I. R., Dopita, M. A., Mathewson, D. S., 
Long, K. S., Szymkowiak, A. E., Culhane, J. L. 1982, ApJ, 255, 440
\reference{} Deeter, J. E., Nagase, F., Boynton, P. E. 1999, ApJ, 512, 300
\reference{} Finley, J. P., Oegelman, H., Hasinger, G., Truemper, J. 1993 ApJ, 
410, 323
\reference{} Jones, T. W. et al. 1998, PASP, 110, 125
\reference{} Leahy, D. A., 1987, A\&A, 180, 275
\reference{} Manchester, R. N., Staveley-Smith, \& Kesteven, M. J. 1993, ApJ, 
411, 756
\reference{} Murray, S. S., et al., 1997, SPIE, 3114, 11
\reference{} Seward, F., D., Harnden, Jr., F. R., \& Helfand, D. J. 1984 ApJ, 
287, L19
\reference{} Seward, F. D., \& Harnden, Jr., F. R. 1994, ApJ, 421, 581
\reference{} Wang, Q. D. \& Gotthelf, E. V. 1998a, ApJ, 494, 623.
\reference{} Wang, Q. D. \& Gotthelf, E. V. 1998b, ApJL, 509, 109
\reference{} Weisskopf, M. C. O'Dell, S. L., van Speybroeck, L. P. 1996, Proc. 
SPIE 2805, Multilayer and Gazing Incidence X-ray/EUV Optics III, 2.
\reference{} Weiler, K. W. \&\ Sramek, R. A. 1988, ARAA, 25, 295

\end{references}
\end{document}